\newbox\rotbox

\documentstyle[sprocl,epsf,rotate]{article}
\begin{document}

\title{LIGHT HADRON MASSES ON COARSE LATTICES WITH IMPROVED ACTIONS}

\author{FRANK X. LEE,$^{a}\,$ DEREK B. LEINWEBER$\,{}^{b}$}
\address{
\it ${}^a$TRIUMF, 4004 Wesbrook Mall,
Vancouver, British Columbia, Canada V6T 2A3 \\
\it ${}^b$Department of Physics, University of Washington,
Seattle, WA  98195}

\maketitle

The advent of mean-field improved perturbation theory \cite{lepage93}
has generated renewed interest in the on-shell improvement of lattice
QCD actions.  The remarkable success of the ${\cal O}(a^2)$-improved
gauge-field action \cite{alford95} has led us to reconsider the
two-link ${\cal O}(a^2)$-improved fermion action of Hamber and
Wu,\cite{hamber83} with mean-field improved operators.

   Tree-level ${\cal O}(a^2)$-improved mean-field improved gauge
configurations are generated in the quenched approximation via the
Cabibbo-Marinari pseudo-heat-bath algorithm.  
%
%
Fermion propagators are obtained via the Stabilized Biconjugate
Gradient algorithm.\cite{frommer94} 
%
%
We consider a local source and both local and gauge-invariant Gaussian
smeared sinks.\cite{gusken90} We found smearing is necessary in order
to extract observables prior to a loss of signal.  Five quark masses
ranging from the strange quark mass to one-third the strange quark
mass are considered.  The strange quark mass was taken to be
approximately 180 MeV.

   To evaluate scaling in the observables, two lattice spacings $a$ are
considered on equal physical volume lattices.  The lattice parameters
are summarized in table \ref{param}.  
%
%
A single-elimination jackknife estimate of the covariance matrix
provides a correlated $\chi^2$ which is used to select the time-slice
analysis regime.

\begin{table}[b]
\vspace{-18pt}
\begin{center}
\caption{Summary of lattice parameters. }
\label{param}
\begin{tabular}{|cccccc|}
\hline
Volume  &$\beta$ &$a_{\rm st}$ &$N_U$ &$u_0$ &$\kappa_{\rm cr}$ \\
\hline
$6^3 \times 12$  &6.25 & 0.40 fm &90 &0.820 &0.1967(2) \\ 
$10^3 \times 16$ &7.00 & 0.24 fm &37 &0.866 &0.1823(3) \\ 
\hline
\end{tabular}
\end{center}
\vspace{-6pt}
\end{table}

Dispersion is examined by calculating the speed of light squared from
the relation $c^2 = (E_p^2 - M^2) / p^2$, to be compared with 1. $E_p$
and $M$ are extracted from lattice correlation functions and $p$ is
the smallest nontrivial momentum $(1,0,0)\, 2\pi/L$.  Table \ref{c^2}
compares $\rho$-meson results for $m_q
\sim 125$ MeV with the ${\cal O}(a)$-improved SW
action\cite{collins95} and the ${\cal O}(a^2)$-improved D234
action\cite{alford95D234}.  The last column tests rotational symmetry
for the Hamber Wu action (HW) by determining $c^2$ from the smallest
two-dimensional diagonal momentum.  The ${\cal O}(a^2)$-improved
actions perform well at the spacing of 0.24 fm.

\begin{figure}[h]
\begin{center}
\epsfysize=8.8truecm
\leavevmode
\setbox\rotbox=\vbox{\epsfbox{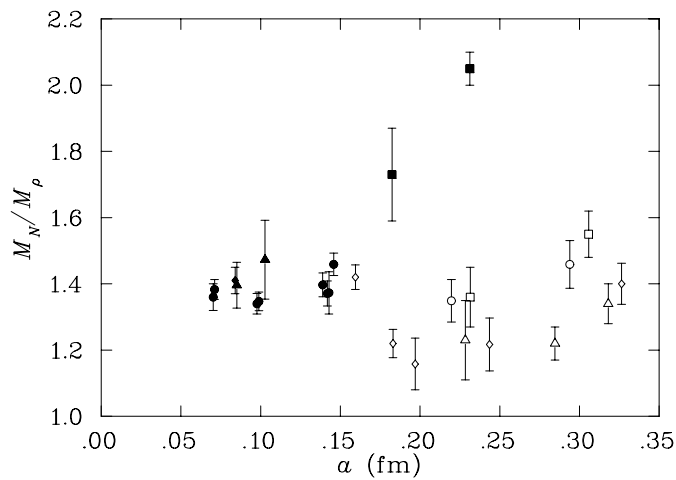}}\rotl\rotbox
\end{center}
\caption{The $N/\rho$ mass ratio as a function of the lattice spacing
$a_\rho$ for a variety of lattice QCD actions.  Solid symbols denote
the standard Wilson action.  Open symbols denote improved actions
including SW \protect\cite{collins95} $(\diamond)$, D234
\protect\cite{alford95D234} $(\triangle)$, HW \protect\cite{fiebig96}
$(\Box)$, and HW $(\circ)$ (this work).\hfill\null} 
\label{MnMrho}
\end{figure}

Hadron mass ratios within the baryon octet, decuplet and the vector
meson nonet scale well and look very similar to experiment.  However,
mass ratios of hadrons with differing spins reveal less than
satisfactory scaling for the larger lattice spacing.  The $N/\rho$
mass ratio summarized in figure \ref{MnMrho} is the most revealing
ratio.  The Hamber and Wu action has reproduced the state of the art
quenched QCD ratio on a tiny $10^3 \times 16$ lattice.  This result
bodes well for future explorations of hadron phenomenology beyond the
quenched approximation.

\begin{table}[t]
\begin{center}
\caption{Evaluation of dispersion and rotational invariance via $c^2$. }
\label{c^2}
\begin{tabular}{|c|cccc|}
\hline
$a_{\rm st}$ &SW (1,0,0) &D234 (1,0,0) &HW (1,0,0) &HW (1,1,0) \\
\hline
0.40         &0.48(4)    &0.93(3)      &0.88(6)    &0.81(9)   \\
0.24         &$--$       &1.00(6)      &0.96(9)    &1.03(7)    \\
\hline
\end{tabular}
\end{center}
\vspace{-12pt}
\end{table}

This work is supported in part by the Natural Sciences and
Engineering Research Council of Canada and U.S. DOE under 
grant DE-FG06-88ER40427.


\begin{thebibliography}{1}

\bibitem{lepage93}
G.~P. Lepage and P.~B. Mackenzie, {Phys.\ Rev.\ D} {\bf 48},  2250  (1993).

\bibitem{alford95}
M. Alford {\it et~al.}, {Phys.\ Lett.} {\bf B361},  87  (1995).

\bibitem{hamber83}
W. Hamber and C. Wu, {Phys.\ Lett.} {\bf 133B},  351  (1983).

\bibitem{frommer94}
A. Frommer {\it et~al.}, {J.\ Mod.\ Phys.} {\bf C5},  1073  (1994).

\bibitem{gusken90}
S. {G\"usken}, {Nucl.\ Phys.\ B} ({Proc.\ Suppl.}) {\bf 17},  361  (1990).

\bibitem{collins95}
S. Collins, {\it et~al.}, {Nucl.\ Phys.\ B} ({Proc.\ Suppl.}) {\bf
47},  366  (1995). 

\bibitem{alford95D234}
M. Alford, {\it et~al.}, {Nucl.\ Phys.\ B} ({Proc.\ Suppl.})
  {\bf 47},  370  (1995).

\bibitem{fiebig96}
H.~R. Fiebig and R.~M. Woloshyn, TRI-PP-96-1, hep-lat/9603001. 

\end{thebibliography}

\end{document}